# SLAC Linac RF Performance for LCLS[*]

R. Akre, V. Bharadwaj, P. Emma, P. Krejcik, SLAC, Stanford, CA 94025, USA

*Abstract*

The Linac Coherent Light Source (LCLS) project at SLAC uses a dense 15 GeV electron beam passing through a long undulator to generate extremely bright x-rays at 1.5 angstroms. The project requires electron bunches with a nominal peak current of 3.5kA and bunch lengths of 0.020mm (70fs). The bunch compression techniques used to achieve the high brightness impose challenging tolerances on the accelerator RF phase and amplitude. The results of measurements on the existing SLAC linac RF phase and amplitude stability are summarised and improvements needed to meet the LCLS tolerances are discussed.

## 1 LCLS RF REQUIREMENTS

LCLS requires the SLAC linac to perform with tolerances on RF phase and amplitude stability which are beyond all previous requirements. The LCLS is divided into four linacs L0, L1, L2, and L3 [1]. The phase and amplitude tolerances for the four linacs operated at S-Band, 2856MHz, are given in Table 1.

Table 1: LCLS RF stability requirements.

|    | Klystrons | Phase rms °S[†] | Amp. % rms |
|----|-----------|-----------------|------------|
| L0 | 2         | 0.5             | 0.06       |
| L1 | 1         | 0.1             | 0.06       |
| L2 | 34        | 0.1             | 0.15       |
| L3 | 45        | 2.0             | 0.05       |

L0 is a new section of accelerator for the off axis injector. L1, L2, and L3 are made of structures in the existing linac from sector 21 to sector 30.

## 2 LINAC RF SYSTEM

### 2.1 The RF Distribution System

The RF distribution and control systems for the linac, after upgrades 15 years ago[2] for the SLAC Linear Collider(SLC) are shown in figure 1. The RF distribution system consists of coaxial lines with varying degrees of temperature stabilisation, figure 1. The 3.125 inch rigid coax Main Drive Line, MDL, carries 476MHz down the 2 miles of accelerator. At the beginning of each of the 30 sectors the 476MHz is picked off and multiplied by 6 to get 2856MHz. There is a temperature-stabilised coaxial Phase Reference Line, PRL, that carries the reference signal to the Phase and Amplitude Detector, PAD, of the eight klystrons in the sector.

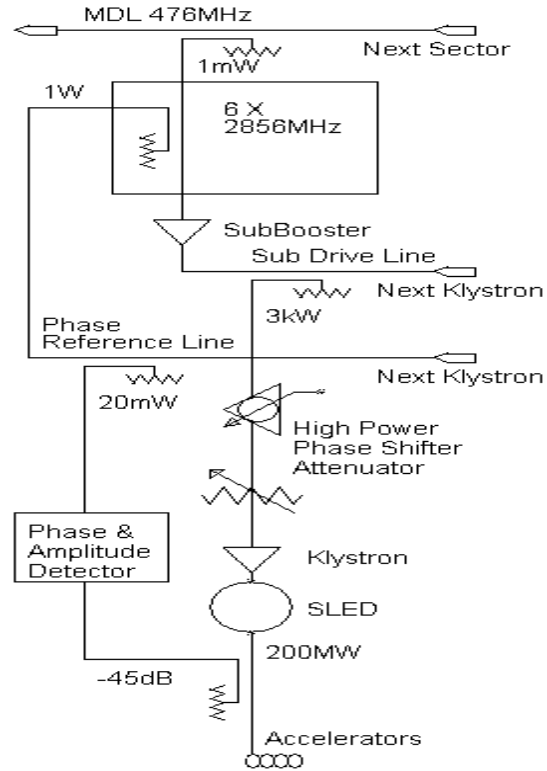

Figure 1: SLAC linac RF station

The critical parameters for the short term and long term variations in the RF phase and amplitude can be read back through the existing control system. The phase and amplitude from the output of the SLED energy storage cavity are compared and recorded by the PAD. There are three methods of acquiring and displaying the data:

- The fast time plot gives 64 consecutive data points. At 30Hz this is 2.1 seconds of data.
- The correlation plot collects data with a maximum frequency of about 1Hz and can collect up to 512 data points.
- The history buffers are updated with a data point every six minutes for the past week and every four hours for the past 7 years.

The bit resolution of the ADC in the PAD is 0.04°S. Phase and amplitude stability has been measured for the different time scales.

---

[*] Supported by the U.S. Department of Energy, contract DE-AC03-76SF00515: LINAC2000 THC11: SLAC-PUB-8574
[†] Throughout this paper °S, °F, and °C stand for degrees at 2856MHz, S-Band, degrees Fahrenheit, and Celsius respectively

## 2.2 RF Phase Stability

Phase fast time plots have an rms variation of 0.05°S and meet LCLS requirements on a two second time scale. On a larger time scale drifts of well over 0.1°S are observed as temperature of the regulating water and environment changes, figure 2. The phase correlation, 6°S/°F, is likely due to the high Q SLED cavity.

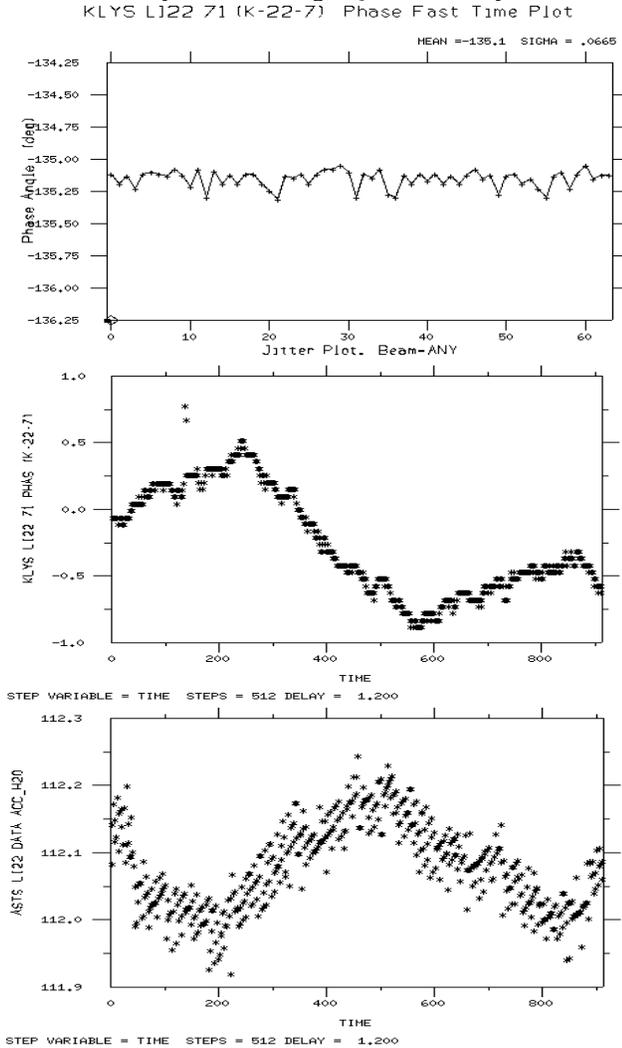

Figure 2: Top: Klystron phase, 2.1-second time scale. Center Klystron phase 14-minute time scale. Bottom SLED water temperature °F 14-minute time scale.

During normal linac operation each klystron's phase is adjusted by a high power phase shifter to keep the phase as read by the PAD within a few degrees of the set value. The phase shifter is a rotary drum type and typically moves about a dozen times a day by a stepper motor. The resolution of the phase shifter is 0.125°S, which is much coarser than the short term phase variation seen on the fast time plots. The position of this phase shifter is recorded in history buffers. The phase shifter movement over a three-day period has been correlated to outside temperature and the coefficients listed in Table 2. The klystrons are grouped according to their position within the sector and averaged over the 29 sectors from sector 2 to sector 30. Position 1 is closest to the sub-booster klystron and position 8 is at the end of the sector. During the course of a year the outside temperature varies from 35°F to 95°F and as much as 35°F diurnally.

Table 2: Klystron phase shifter movement

| Klystron Position | Average °S/°F | Standard Deviation | Range °S |
|---|---|---|---|
| 1 | 0.33 | 0.11 | 20 |
| 2 | 0.41 | 0.10 | 25 |
| 3 | 0.46 | 0.11 | 28 |
| 4 | 0.49 | 0.14 | 29 |
| 5 | 0.60 | 0.14 | 36 |
| 6 | 0.69 | 0.13 | 41 |
| 7 | 0.80 | 0.16 | 48 |
| 8 | 0.64 | 0.19 | 38 |

## 2.3 RF Phase Measurement Accuracy

The critical phase stability of the RF with respect to the beam is influenced at three levels within the RF distribution and control system. The first level is the stability of the phase reference system. The second tier is the noise level and drifts associated with the phase measurement electronics, and the third level consists of the errors introduced in the beam phase measurement system.

The two-mile MDL has been studied [3] and the length electronically measured by an interferometer. From reference [3] the length varies with pressure and temperature over the 2 miles as follows:

$$\Delta\phi(°S) \sim -2.64(\Delta P(mBar)) + 1.36(\Delta T(°F))$$

History buffers show that the pressure range, $\Delta P$, is about 30mBar, which gives a phase variation of 79°S. The temperature range, $\Delta T$, of the MDL is about 30°F, half the outside $\Delta T$ due to some insulation and temperature regulation. This $\Delta T$ gives a phase variation of 41°S. The predicted phase variations based on the above analysis only accounts for about half the observed phase tuning in the linac that is necessary to keep the beam at constant phase to meet the beam energy and energy spread requirements [4]. These additional errors indicate the system is in need of an upgrade.

About 95% of the PRL is temperature controlled with an rms value of 0.05°F. The other 5% varies by about 10% of the surrounding temperature, which gives a temperature variation of about 1.0°F. The ½ inch heliax has a temperature coefficient of 4ppm/°C, 0.9°S/°F/sector. The phase error is spread linearly from a minimum at the first klystron in the sector to a maximum at the eighth klystron in the sector. The average phase variation of the sector is ½ the phase variation of the PRL, 0.5°S.

The multipliers are temperature stabilised to about 0.1°Frms and have temperature coefficients which range from –1.7°S/°F to +2.2°S/°F. The phase errors from the multipliers are on the order of 0.2°S rms.

Additional errors are introduced between the phase reference system and the beam by the variations in length due to temperature of the accelerating sections and the waveguide feeding them. These variations are ignored by the feedback system since the PAD only measures the signal at the output of the SLED cavity. Table 3 summarises the phase errors due to temperature changes in the system. The dominant non-corrected error is due to the accelerator structure temperature change. Measuring the RF phase at the output or input of the structure as an estimate for the phase of the structure as seen by the beam would have an error of 0.8°S rms, or half the phase slippage of the structure.

Table 3: Phase/Temperature coefficients

|  | °S/°C | ΔTrms °C | Δϕrms °S |
|---|---|---|---|
| Accelerator 10' [5] | 16.0 | 0.1 | 1.6 |
| WR284 Cu WG 10' [5] | 0.25 | 0.2 | 0.05 |
| ½" Heliax 40' @ 4ppm/°C | 0.16 | 1.0 | 0.16 |
| 7/8" Heliax 40' @ 3ppm/°C | 0.12 | 1.0 | 0.12 |
| 1-5/8" Rigid 40' MDL data | 0.01 | 0.1 | 0.001 |
| SLED [6] | 23.6 | 0.1 | 2.4 |
| PAD | <0.5 | 0.1 | <0.05 |

The measurement resolution of the PAD is good enough to meet the LCLS requirements. Initial testing show measurement drifts of the PAD from temperature variations to be close to LCLS requirements. Further testing will be done to better estimate the PAD errors.

## 2.4 RF Amplitude Stability

Fast time plots for the klystron amplitude also show that on a two second time scale the LCLS stability criterion can be met. The rms amplitude jitter measured by the PAD at the output of he SLED cavity is less than 0.04% of the amplitude. Correlation plots over a 14-minute time scale show the amplitude varies by as much as 0.5% peak to peak. This change is correlated to the water temperature of the SLED cavity and the magnitude of variations is greatly effected by the tune of the cavity[7]. Klystron K02 on the SLAC accelerator has a slow amplitude feedback and no SLED cavity. Measurement of the amplitude variation over days is held to 0.06% rms. Further work needs to be completed to determine how stable the measurement is with respect to temperature changes.

## 3 RF SYSTEM IMPROVMENTS

Extremely tight phase and amplitude tolerances throughout the linac are required to meet the LCLS specifications. The LCLS requirements listed in Table 1 may still change as the design of the bunch compression system evolves. Measurement of the individual klystrons show that they are capable of attaining the desired specification up to a two second time scale. The challenge is to link the many klystrons together through a RF distribution system and preserve the stability over extended periods of time.

On longer time scales where temperature changes are significant, a new RF reference and distribution system located in the tunnel, which has rms temperature variations less than 0.1°F, is under consideration. The new system will distribute 2856MHz to the klystrons and provide a reference for phase measurements of the accelerator RF and beam phase cavity RF. This new phase system is expected to reduce the phase drifts and errors along the kilometer linac from about 10°S down to as little as 0.1°S. Even with such a phase stable RF reference system, measuring the phase of the RF at the input or output of the accelerator will result in errors of 0.8°Srms compared to the RF phase as seen by the beam in an accelerator structure which has temperature variations of 0.1°Crms. In order to hold the RF to beam phase to 0.1°S a feedback system using a beam-based measurement is necessary.

Further measurements will determine if the existing amplitude measurement and control system with added feedback is sufficient to meet LCLS requirements.

In LCLS L2 and L3, where there is a large number of klystrons, it is likely that the phase errors will be correlated with water temperature which spans groups of 16 klystrons, or outside temperature and pressure, which is common to all. The larger number of klystrons does not increase the tolerance of an individual klystron by √n.

Further testing of the existing RF system as well as development and testing of new systems is ongoing, the results of which will lead to the design of the LCLS RF system.